# Twisted Nano-optics: Manipulating Light at the Nanoscale with Twisted Phonon Polaritonic Slabs

*Jiahua Duan[1,2], Nathaniel Capote-Robayna[3], Javier Taboada-Gutiérrez[1,2], Gonzalo Álvarez-Pérez[1,2], Iván Prieto[4], Javier Martín-Sánchez[1,2], Alexey Y. Nikitin[3,5]\*, and Pablo Alonso-González[1,2]\**

1 Department of Physics, University of Oviedo, Oviedo 33006, Spain.

2 Center of Research on Nanomaterials and Nanotechnology, CINN (CSIC-Universidad de Oviedo), El Entrego 33940, Spain.

3 Donostia International Physics Center (DIPC), Donostia/San Sebastián 20018, Spain

4 Institute of Science and Technology Austria, am Campus 1, Klosterneuburg 3400, Austria

5 IKERBASQUE, Basque Foundation for Science, Bilbao 48013, Spain

**Abstract**

Recent discoveries have shown that when two layers of van der Waals (vdW) materials are superimposed with a relative twist angle between them, the electronic properties of the coupled system can be dramatically altered. Here, we demonstrate that a similar concept can be extended to the optics realm, particularly to propagating polaritons – hybrid light-matter interactions –. To do this, we fabricate stacks composed of two twisted slabs of a vdW crystal ($\alpha$-MoO$_3$) supporting anisotropic phonon polaritons (PhPs), and image the propagation of the latter when launched by localized sources. Our images reveal that, under a critical angle, the PhPs isofrequency curve undergoes a topological transition, in which the propagation of PhPs is strongly guided (canalization regime) along predetermined directions without geometric spreading. These results demonstrate a new degree of freedom (twist angle) for controlling the propagation of polaritons at the nanoscale with potential for nano-imaging, (bio)-sensing or heat management.

.

KEYWORDS: Light canalization, Phonon Polaritons, van der Waals materials, s-SNOM, nano-optics, hyperbolic materials.

The discovery of flat-band superconductivity[1, 2, 3] and ferromagnetism[4, 5] in twisted bilayer graphene has boosted an intense research in twisted layered structures composed of vdW crystals, aiming to uncover novel physical phenomena as a function of a relative twist angle ($\theta$) between the in-plane crystalline axes of individual layers[1, 6, 7, 8, 9]. In particular, twisted layered structures have already impacted optics and photonics, by the recent demonstrations of atomic photonic crystals in twisted bilayer graphene[10], moiré excitons in stacked structures composed of transition-metal dichalcogenides[9, 11, 12, 13, 14], and plasmons in graphene moiré superlattices[15] and magic-angle graphene bilayers[16]. Interestingly, two recent theoretical reports have predicted that a rich plethora of novel polaritonic phenomena, including topological transitions of polariton dispersions and broadband field canalization, also arises within twisted hyperbolic (meaning that a slice of the polariton dispersion surface by a plane of constant frequency, the so-called isofrequency curve or IFC, defines a hyperbolic curve) metasurfaces via near-field coupling of polaritons[17,18]. In this regard, the recent demonstration of infrared phonon polaritons (PhPs)[19] –light coupled to lattice vibrations– exhibiting in-plane anisotropic propagation with ultra-low losses and sub-diffractional confinement in the vdW polar crystal α-MoO$_3$[20, 21], provides an optimal test bench to study near-field mediated coupling effects on the propagation of PhPs in twisted layered structures.

Here, we experimentally demonstrate effective manipulation of the propagation direction of PhPs on the nanoscale by making use of vdW twisted structures. To do so, we stack two α-MoO$_3$ slabs with a relative rotation angle between their in-plane optical axes and study the propagation of PhPs along them by infrared nanoimaging. Remarkably, we uncover a topological transition of the polaritonic IFC from an open hyperbola-like curve to a closed circumference-like one, finding a critical angle where the IFC flattens, giving rise to strongly directional and diffraction-less propagation of PhPs –canalization regime– along one specific in-plane direction. We corroborate our experimental results by full-wave simulations as well as by simple analytical modelling, developed to easily represent the IFCs of twisted bilayer polaritonic structures.

To study the propagation of PhPs in twisted bilayers of α-MoO$_3$, we first perform full-wave numerical simulations, in which we model two superimposed slabs of α-MoO$_3$ with identical thickness d = 200 nm and whose in-plane crystallographic axes are mutually rotated by an angle θ (Figure 1a). To excite PhPs in the stacked structure, we place a vertically-oriented electric dipole (at an illuminating wavelength $\lambda_0$ = 11.0 μm) close to the top surface (see Methods), generating highly confined electromagnetic fields that provide large momenta. We start our numerical study for the case θ = 0º , i.e. when the respective crystalline axes of both slabs are aligned. Figure 1b (left) shows the real part of the electric field along the z direction, Re($E_z$(x, y)), revealing the fundamental PhPs mode in the biaxial slab[22] propagating with concave wavefronts centered along the α-MoO$_3$ [100] crystal direction. The corresponding IFC in **k**-space (with **k** = ($k_x$, $k_y$) the two-dimensional (2D) in-plane wavevector), is visually represented by Fourier transforming (FT) the the field $E_z$(x, y). The IFC presents an open hyperbola-like curve with the transverse and conjugate axes coinciding with the [100] and [001] crystal axes, respectively, showing a range of forbidden directions with no available *k* along them. Note that to better correlate the real-space plots and IFCs, it is helpful to consider the Poynting vector **S**, as it determines the energy propagation direction of PhPs, being normal to the IFC (see arrows in the figure). This in-plane hyperbolic propagation of PhPs is qualitatively similar to that of single slabs of α-MoO$_3$ as previously reported[20, 21]. Therefore, as expected, a stacked structure with θ = 0º does not introduce any change in the propagation direction of PhPs in α-MoO$_3$ since the crystallographic axes (and thus the optical axes) are the same. At θ = 30º we observe again PhPs propagating with concave wavefronts (Figure 1c); however, they appear now centered along a direction that is tilted at an angle φ = 12º with respect to the [100] crystal axis of the top α-MoO$_3$ layer. The tilting effect is also visualized in the corresponding IFC, where an open hyperbola whose transverse axis forms an angle φ with the [100] crystal axis is clearly seen. The change of the PhPs propagation angles by φ is attributed to polaritonic coupling between the slabs in the twisted stack, giving rise to novel polaritonic propagation regimes, in comparison to single slabs. A particularly remarkable propagation regime occurs at θ=63º (Figure 1d), where PhPs seem to propagate only along one specific in-plane direction at φ=25º Accordingly, the IFC appears flattened and shows a unique direction for **S** at φ=25º uncovering the so-called canalization regime in the stacked structure, which holds great promises for guiding and manipulating the energy flow of electromagnetic fields at the nanoscale[23]. Intriguingly, when we further increase the twist angle to θ = 90º, PhPs are found to propagate along all in-plane directions (Figure 1e), as described by a closed IFC. Thus, our simulations show that the topology of the polaritonic IFC can thus be tuned in twisted slabs from an open hyperbola to a closed shape when the twist angle θ varies from 0 to 90º. In particular, they reveal a polaritonic topological transition at a critical angle θ = 63º, where the IFC flattens. These results point out that the propagation of PhPs can be effectively engineered by considering stacks of twisted slabs of α-MoO$_3$. Together with the FTs of the simulated electric field distribution, we also plot the analytical IFCs (green curves in Fig. 1b-e), calculated upon a simplified model based on the small thickness of both layers, thus not requiring to compute the fields inside the slabs (see Supporting Information). Our approximation is similar to the approximations developed in references[17,18], but valid for two twisted biaxial layers of a finite thickness and separated by an arbitrary distance. A good agreement between the results of the simulations and the model allows us to use the latter for a quick qualitative estimation of the IFC.

Based on our theoretical findings, in the following, we experimentally demonstrate the modification of the IFCs and the emergence of a canalization regime for PhPs in twisted α-MoO$_3$ slabs. To that end, we fabricate stacks of two α-MoO$_3$ slabs with different relative twist angles θ (see Methods), and visualize the propagation of PhPs launched along them by metal antennas (acting as localized sources) by infrared nanoimaging employing a scattering-type near-field optical microscope[24,25] (s-SNOM, see Methods). As schematized in Figure 2a, the s-SNOM tip and sample are illuminated with s-polarized mid-infrared light —with the electric field parallel to the longitudinal direction of the metal antenna—, spanning wavelengths from 10.8 to 11.2 μm within the hyperbolic Reststrahlen band of α-MoO$_3$[26] (RB$_2$: 10.4 - 12.2 μm). Importantly, due to the much larger launching efficiency of the resonant gold antenna compared to the s-SNOM tip under such illumination scheme, direct visualization of the PhPs wavefronts is possible. For a twist angle θ = 30º, we observe that the polaritonic wavefronts are concave and propagate along a direction φ ~ 10º with respect to the [100] crystal axis of the top α-MoO$_3$ slab (Figure 2b). The experimental IFC (left bottom panel), extracted from the FT of the near-field image, shows an open hyperbola with its transverse axis tilted at an angle φ with respect to the [100] crystal axes of the top α-MoO$_3$ layer, in excellent agreement with our numerical simulations mimicking the experiment (the FT of the simulated electric field distribution is shown in the right bottom panel, see Methods). Excitingly, at θ = 54º, the propagation of PhPs appears to be strongly guided along one specific direction along φ = 28º (Figure 2c), yielding a flattened IFC (left bottom panel) corresponding to a canalization regime, again in excellent agreement with numerical simulations (right bottom panel). We note that there is, however, a small difference between the critical angles from the experiment (θ$_c$ = 54º) and the prediction in Figure 1, (θ$_c$ = 63º), that we attribute to a slightly different thickness for the top and bottom α-MoO$_3$ layers in our fabricated stacks, as it is challenging to exfoliate several homogenous α-MoO$_3$ flakes with exactly the same thickness. When θ = 85º, we observe PhPs propagating along all in-plane directions, yielding a closed IFC (Figure 2d), which is also observed for θ = 90º (Figure 2e). The IFCs of these two cases are, however, slightly different, being the IFC of PhPs for θ = 90º more squared-like, pointing out that even a small difference of 5º on the twist angle significantly alters the propagation of PhPs, which evidences twisting as a rather sensitive method for the manipulation of the polaritons propagation. In agreement with the predictions shown in Figure 1, our experimental results demonstrate that: (i) a topological transition from an open hyperbola (Figure 2b) to a closed IFC (Figure 2e) takes place as a function of the twist angle θ, and (ii) canalization of PhPs occurs at specific twist (critical) angles θ$_c$.

Given the strong dispersive nature of PhPs in α-MoO$_3$[20, 21], we also explore their propagation in a twisted structure (at θ = 90º) as a function of incident wavelength λ$_0$ from 10.8 μm to 11.1 μm (Figure 3). For λ$_0$ = 10.8 μm, near-field imaging shows PhPs propagating with concave wavefronts and within hyperbolic sectors whose transverse axes lay along both the [100] and the [001] crystal directions (Figure 3a), appearing to be less pronounced along the latter. We explain this difference by interference of PhPs launched at different antenna locations along the [100] direction, which makes the polaritonic pattern to flatten. The corresponding IFC consists of a hyperbola with its transverse axis along the [100] crystal axis of the top α-MoO$_3$ layer and two bright spots along the [001] crystal axis (bottom left panel). This IFC agrees well with our numerical simulations (bottom right panel), showing an open hyperbola along the two in-plane crystal directions. Interestingly, these two hyperbolas do not intersect, so that the topology of the IFC in each slab is not affected by the stacking, revealing the absence of polaritonic coupling between the slabs and, therefore, the propagation of PhPs is independent in each of them. For λ$_0$ = 10.9 μm, we observe again PhPs propagating within hyperbolic sectors centered along both in-plane crystal axes; however, intriguingly, we also observe PhPs along previously forbidden directions (Figure 3b), which indicates that at this wavelength, due to coupling of the PhPs in the upper and lower slabs, the IFC topology changes, giving rise to a closed IFC curve, both in experiment and theory (bottom panels). This coupling effect is even more remarkable for λ$_0$ = 11.0 μm, where we clearly visualize PhPs propagating along all in-plane directions (Figure 3c), and thus the polaritonic IFC forms an entirely square-like closed shape in both experiment and theory (bottom panels). Similarly, PhPs propagate along all in-plane directions for λ$_0$ = 11.1 μm (Figure 3d), yet in a slightly different manner, as the IFC evolves from a square-like to a circumference-like closed shape, which uncovers nearly isotropic propagation of PhPs. These near-field measurements demonstrate that the propagation of PhPs in a twisted polaritonic structure can also be conveniently tuned by varying the incident wavelength, as a result of wavelength-dependent polaritonic coupling between the slabs, that, excitingly, also induces a topological transition of polaritonic IFCs from open to closed curves.

Canalization of PhPs —that is, propagation of PhPs along one specific direction $\varphi$— offers unique possibilities for nanoscale guiding of electromagnetic fields and can occur in our twisted structures at certain critical twist angles $\theta_c$, as shown in Figure 2c. To get more insights into the propagation of PhPs in the canalization regime of our stacked α-MoO$_3$ slabs, we extract a line-profile from the near-field amplitude image taken at $\lambda_0 = 11$ μm and $\theta_c = 54°$ (dashed green line in Figure 4a), and fit it to an exponentially-decaying oscillating function which takes into account both the polaritonic intrinsic damping and its geometrical decay[27], having the following form:

$$\xi(x) = \frac{A}{x^f} e^{-ik_x x}, \quad A, k_x, f > 0, \tag{1}$$

where $k_x$ is the complex polariton wavevector and $f$ the geometrical decay factor. Interestingly, the resulting curve (green line in the left panel of Figure 4b) yields a value of 0 for $f$, meaning that the geometrical decay in the canalization regime is negligible, and thus the PhPs propagate diffraction-less, in stark contrast to the case of a single α-MoO$_3$ slab (orange curve), where $f$ takes a value of 0.5, in agreement with previous reports[20,26,28]. This result is further corroborated by numerical simulations mimicking the experiments (right panel in Figure 4b). To better exploit the potential application of such diffraction-less propagation of PhPs in our stacked structures at critical angles, we numerically study the expected values of $\theta_c$ and propagation angles $\varphi$ as a function of the incident wavelength $\lambda_0$ (Figures 4c, and 4d, respectively, where stacked structures consisting of two rotated α-MoO$_3$ slabs with the same thickness ranging from 50 to 700 nm, are considered). Interestingly, in all cases we observe a large variation of $\theta_c$ from ~10° to ~70° (Symbols in Figure 4c. Red dots indicate the case for a typical thickness of 200nm), in the rather narrow spectral range from 10.8 μm to 11.6 μm, which can be explained by the strongly dispersive nature of in-plane hyperbolic PhPs in α-MoO$_3$. More importantly, the propagation angle φ of "canalized" PhPs also shows in all cases a variation from ~5º to ~30º (Symbols in Figure 4d. Red dots indicate the case for a typical thickness of 200nm) for the same spectral range of incident wavelengths. The similar values obtained for both $θ_c$ and φ despite the large variation of slab thicknesses considered, indicate the robustness of the interlayer coupling in the twisted structures. Note that in this study we have only considered twisting angles from 0 to 90º, as angles from 0 to -90º would result in complementary values ($-θ_c$,-φ) which, however, would extend the range of expected angles by a factor of 2. Together with these numerical simulations, we plot our experimental results (stars in Figures 4c, and 4d), which show a good agreement, being the small discrepancy between them attributed to the slightly different thicknesses of the two slabs in the fabricated stack. Such tunability of *φ*, which in the future could be extended by considering a controlled difference of thicknesses of the slabs, has thus great potential for routing flows of energy at the nanoscale.

In summary, we report unprecedented control of the propagation of PhPs in twisted vdW structures composed of two mutually rotated layers of the hyperbolic crystal α-MoO$_3$. Our experimental observations, acquired by imaging the propagation of PhPs in real space, unveil a topological transition of the polaritonic IFC from open hyperbolas to closed-shaped curves as a function of both the twist angle and the incident wavelength. Excitingly, through this topological transition and under certain conditions (critical angle), a flat-IFC state emerges, giving rise to canalization of PhPs with diffraction-less propagation. Our findings demonstrate that the principles used to control electronic bands in "twistronics" can be analogously extended for controlling polaritons at the nanoscale, setting the grounds for a "twistoptics" field where hyperbolic twisted α-MoO$_3$ structures occupies a privileged site for controlling the flow of light at the nanoscale, the long-pursued goal of nanophotonics.

During the preparation of this manuscript, we became aware of a similar work dealing with the observation of topological polaritons in twisted MoO$_3$ bilayers[29].

AUTHOR INFORMATION

Corresponding Author

*pabloalonso@uniovi.es, alexey@dipc.org.

**Methods.** *Fabrication of twisted α-MoO$_3$ structures.* Twisted stacks of two α-MoO$_3$ slabs were fabricated using the dry transfer technique[30]. To that end, we first performed mechanical exfoliation from commercial α-MoO$_3$ bulk materials (Alfa Aesar) using Nitto tape (Nitto Denko Co., SPV 224P). Afterwards, we performed a second exfoliation of the α-MoO$_3$ flakes from the tape to a transparent polydimethylsiloxane (PDMS) stamp and inspected the transferred flakes with an optical microscope, selecting homogeneous and large pieces with the desired thickness. The PDMS stamp with α-MoO$_3$ flakes on top was mounted in a micromanipulator for alignment and twisting. After the first α-MoO$_3$ flake was peeled off from the PDMS to a SiO$_2$ substrate by heating the sample to 200 ºC, the second α-MoO$_3$ flake was precisely aligned, twisted at the desired angle and released on top of the first one.

*Fabrication of gold antennas.* High-resolution electron beam lithography (100 kV and 100 pA) was carried out with sub-micron alignments on our samples coated with a PMMA resist layer. With conventional high-resolution developer (1:3 MIBK: IPA), evaporation of 5 nm Cr and 30 nm Au and a lift-off were done to define the antennas. In order to remove any organic residual, we performed a hot acetone bath at 60 ºC for 10-15 min and a gentle rinse of IPA for 1 min, followed by a nitrogen gas drying and thermal evaporation. The dimensions of gold antenna were 3 μm (length) x 50 nm (width) x 40 nm (height). We note that α-MoO3 layers are sensitive to any ultrasonic processing.

*Near-field nanoimaging.* All near-field measurements in this paper were performed using a commercial s-SNOM system (Neaspec GmbH) equipped with quantum cascade lasers (Daylight Solutions), in the range 890cm$^{-1}$ – 1140cm$^{-1}$. The s-SNOM is based on an atomic force microscopy (AFM) operating in the tapping mode at $\Omega \sim$ 280 kHz and an amplitude of ~ 100 nm. The AFM tips used were commercial (ARROW-NCPt-50, Nanoworld) and metal-coated (Pt/Ir). The gold antenna was illuminated with s-polarized light from a mid-infrared laser beam, while the tip-scattered light was focused by a parabolic mirror towards an infrared detector (Kolmar Technologies) allowing for direct imaging of the PhPs propagation. A pseudo-heterodyne interferometric method was applied to extract both the amplitude and phase signal and the detected signal was demodulated at the 3$^{rd}$ harmonic to suppress the far-field background scattering.

*Numerical simulations.* Full-wave numerical simulations were performed to calculate the vertical component of the near-field using the finite boundary elements method. The twisted α-MoO$_3$ heterostructures were modelled as two biaxial slabs[26] rotated with respect to each other an angle θ on top of a SiO$_2$ substrate. PhPs were launched by electric sources placed on top of the structure, either (i) a vertically-oriented point dipole (as in Figure 1) or (ii) metal antennas, whose dimensions are the same as the ones used in experiments, and which are illuminated by s-polarized light along their longitudinal axis (as in Figures, 2, 3 and 4). The critical angle (in Figure 4c and 4d) corresponds to the relative orientation of the stacked slabs for which a flattened isofrequency curve is extracted from the simulated near-field distribution of propagating polaritons on the surface of the structure.

Author Contributions

P.A.-G. and A.Y.N. conceived the study with the help of J.D. P.A.-G. and A.Y.N. supervised the project. J.D. fabricated the twisted samples and carried out the near-field imaging experiments with the help of J.T.-G. and J.M.-S. P.A.-G, J.D., G.A.-P, J.T.-G., J.M.-S., A.Y.N. participated in data analysis. N.C.-R. carried out numerical simulations and analytical calculations with the help of G.A.-P. and J.M.-S. I.P. contributed to gold antenna fabrication. J.D, P.A.-G., J.M.-S., G.A.-P, and A.Y.N. contributed to the writing of the manuscript.

**ACKNOWLEDGMENTS**

J.T.-G. and G.Á.-P. acknowledge support through the Severo Ochoa Program from the Government of the Principality of Asturias (nos. PA-18-PF-BP17-126 and PA20-PF-BP19-053, respectively). J. M-S acknowledges financial support through the Ramón y Cajal Program from the Government of Spain (RYC2018-026196-I). A.Y.N. acknowledges the Spanish Ministry of Science, Innovation and Universities (national project no. MAT201788358-C3-3-R). P.A.-G. acknowledges support from the European Research Council under starting grant no. 715496, 2DNANOPTICA.

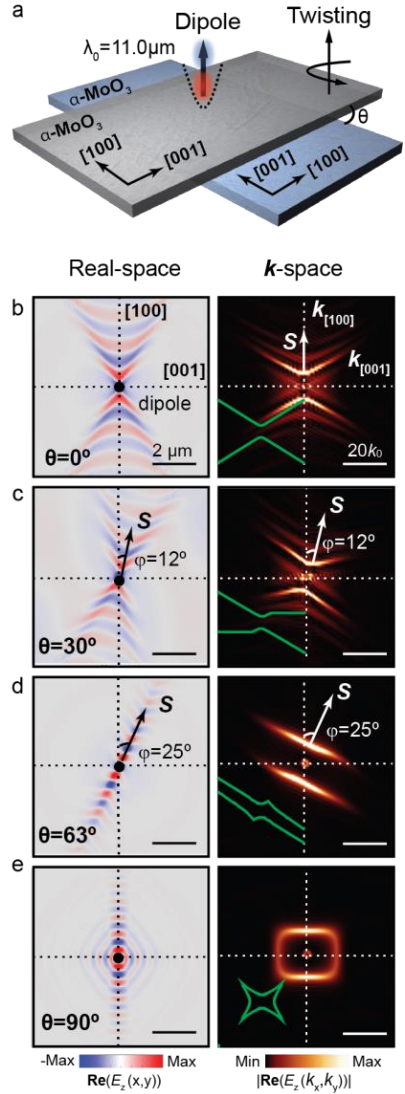

**Figure 1. PhPs in hyperbolic twisted vdW structures. (a)** Schematics of two stacked 200-nm-thick α-MoO$_3$ slabs with their in-plane crystalline directions rotated by a twist angle θ. A vertically-oriented electric dipole, placed on top of the structure, originates highly concentrated electric fields that allow for launching PhPs at an incident wavelength $\lambda_0$ = 11.0 μm. **(b-e)** Simulated real-space field distribution of PhPs propagating in the stacked structure and corresponding polaritonic IFC for twist angles θ=0º (b), 30º (c), 63º (d) and 90º (e), showing a topological transition from an open hyperbola (b) to a closed curve (e), and canalization of PhPs at the critical angle (d). The green solid lines in the insets represent corresponding analytical calculations for IFCs of PhPs in stacked MoO$_3$ slabs.

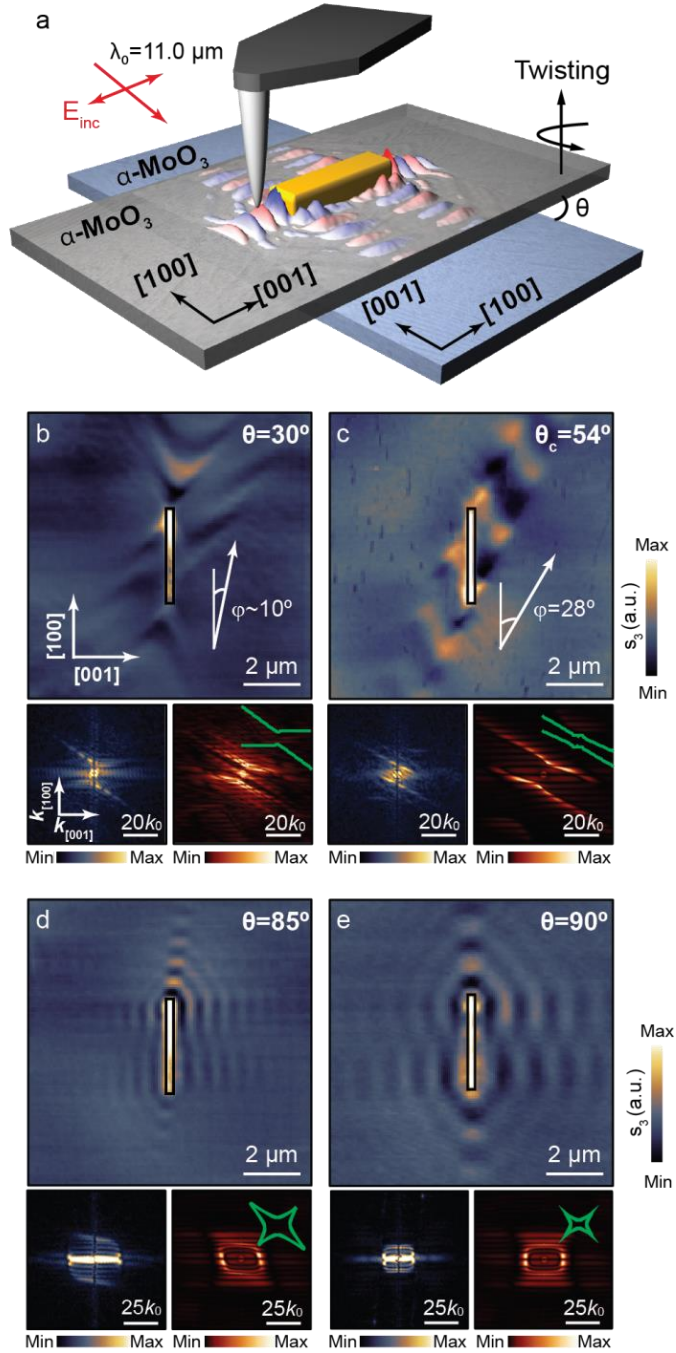

**Figure 2. Real-space imaging of PhPs propagating in twisted α-MoO₃ structures as a function of twist angle *θ*. (a).** Schematics of the s-SNOM experimental scheme used to image PhPs on twisted α-MoO₃ structures. A gold antenna (yellow) confines infrared light ($\lambda_0$=11.0 μm) that allows to launch PhPs, which are probed by a metallized AFM tip. **(b-e)** Near-field amplitude images s₃ (Methods) of twisted α-MoO₃ structures at an illuminating wavelength $\lambda_0$ = 11.0 μm for different rotation angles θ = 30º (b), 54º (c), 85º (d) and 90º (e). The bottom panels show experimental (left) and simulated (right) IFCs for the near-field images in (b-e) and numerical

17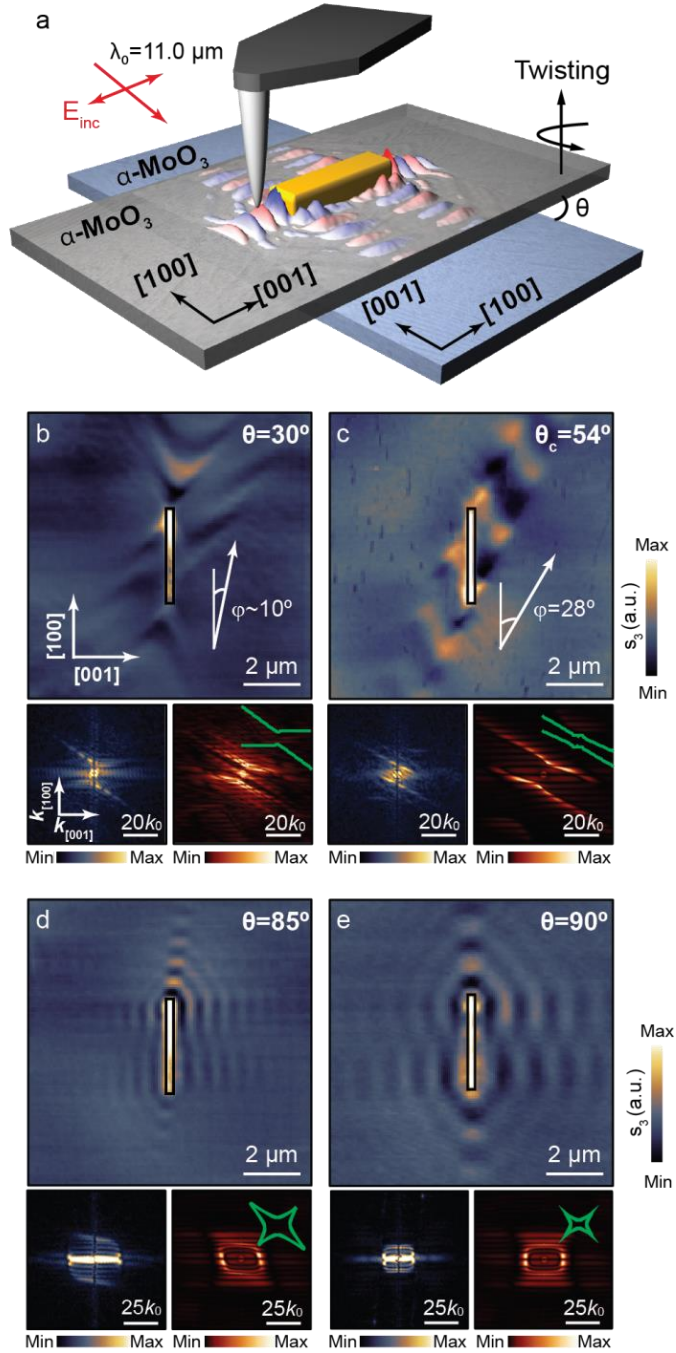

**Figure 2. Real-space imaging of PhPs propagating in twisted α-MoO₃ structures as a function of twist angle *θ*. (a).** Schematics of the s-SNOM experimental scheme used to image PhPs on twisted α-MoO₃ structures. A gold antenna (yellow) confines infrared light ($\lambda_0$=11.0 μm) that allows to launch PhPs, which are probed by a metallized AFM tip. **(b-e)** Near-field amplitude images s₃ (Methods) of twisted α-MoO₃ structures at an illuminating wavelength $\lambda_0$ = 11.0 μm for different rotation angles θ = 30º (b), 54º (c), 85º (d) and 90º (e). The bottom panels show experimental (left) and simulated (right) IFCs for the near-field images in (b-e) and numerical

simulations mimicking the experiment, respectively. A clear topological transition from an open hyperbola (b) to a closed curve (e) as a function of the twist angle is observed, and canalization of PhPs at the critical angle $\theta_c=54º$ (c). The thickness of the top/bottom α-$MoO_3$ layers are 240 nm/180 nm, 180 nm/280 nm, 90 nm/120 nm, 200 nm/200 nm in (b,e), respectively. The corresponding analytical IFCs are plotted as green solid lines. The white arrows in (b-c) represent the Poynting vector, indicating the energy propagating direction of PhPs.

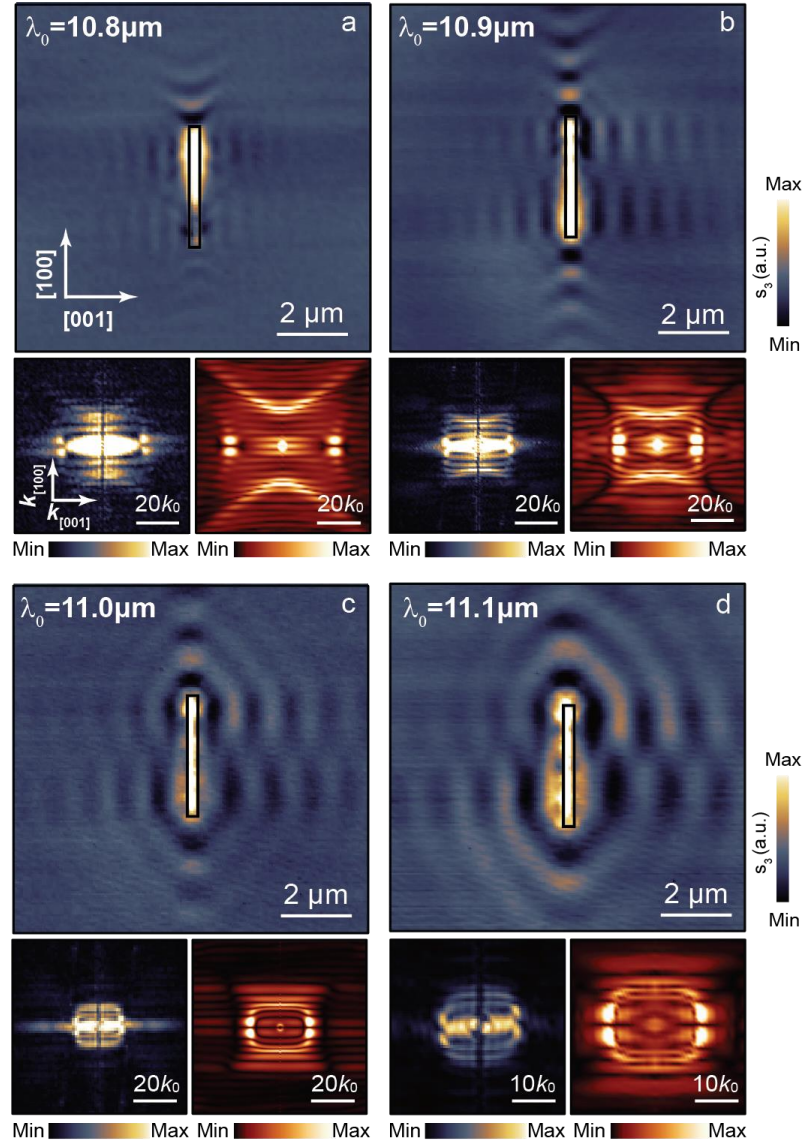

**Figure 3. Real-space imaging of PhPs propagating in 90º-twisted α-MoO₃ stacked structure as a function of incident wavelength $\lambda_0$. (a-d).** Near-field amplitude images s₃ of α-MoO₃ twisted structures for $\lambda_0$ = 10.8 μm (a), $\lambda_0$ = 10.9 μm (b), $\lambda_0$ = 11.0 μm (c), and $\lambda_0$ = 11.1 μm (d). The bottom panels show the corresponding experimental (left) and simulated (right) IFCs for the near-field images in (a-d) and numerical simulations mimicking the experiment, respectively. A clear topological transition from an open hyperbola (a) to a closed curve (d) as a function of the incident wavelength is observed. The thickness of the top/bottom α-MoO₃ layers are 200 nm/200 nm.

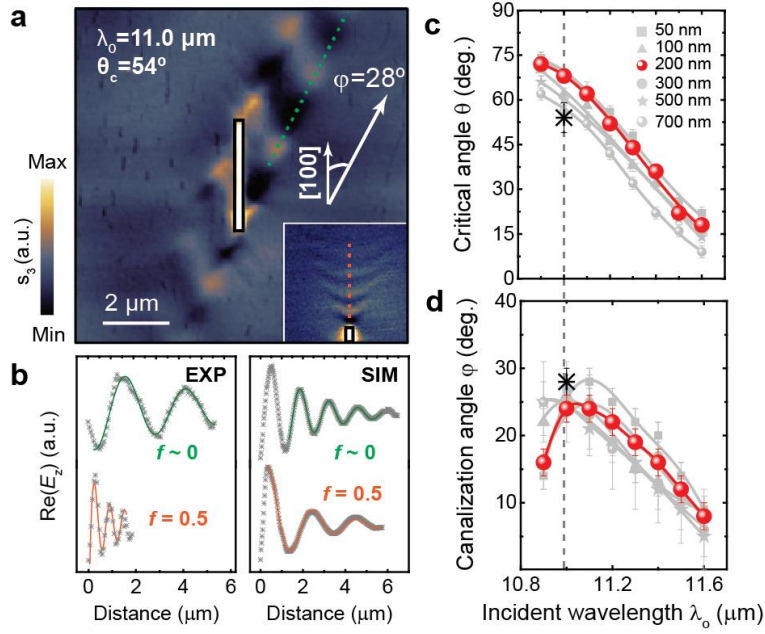

**Figure 4. Canalization of PhPs in twisted α-MoO₃ stacked structures. (a)** Near-field amplitude image s₃ of a α-MoO₃ stacked structure at an illuminating wavelength $\lambda_0 = 11.0$ μm and a twist angle θ = 54º. The thickness of the top/bottom α-MoO₃ layers are 180 nm/280 nm, respectively. For comparison, the inset in the image shows a near-field amplitude image of PhPs propagating on a single slab of α-MoO₃ with a thickness of 100 nm. **(b)** *Left*: line-profiles (gray symbols) along the near-field experimental images in (a) (green and orange dashed lines for the stacked structure and the single slab (bottom), respectively), together with fitted curves following Eq. 1 (green and orange solid lines for the stacked structure and the single slab, respectively). *Right*: line-profiles (gray symbols) along near-field simulations (not shown) for canalization in a stacked structure (top) and hyperbolic propagation in a single slab (bottom), together with fitted curves following Eq. 1 (green and orange solid lines for the stacked structure and the single slab, respectively). **(c-(c-d)** Dependence of the critical angle $\theta_c$ and the propagation direction $\varphi$ on the incident wavelength $\lambda_0$. The thicknesses of the top/bottom α-MoO₃ layers in the simulations are set to 50nm/50nm, 100 nm/100 nm, 200 nm/200 nm, 300 nm/300 nm, 500 nm/500 nm and 700 nm/700 nm. Red dots indicate the case for a typical thickness of 200 nm, while grey points represent other corresponding thicknesses (interpolated curves are also shown for each case). Black symbols represent the experimental results.